# 复杂系统中的量子精密测量及其量子模拟实验检验

艾清 [1]，王洋洋 [2,3]，邱静 [4]

1 北京师范大学物理学系，100875，北京

2 陕西省可控中子源工程技术研究中心，西京学院电子信息学院，710123，西安

3 陕西省可控中子源应用技术国际联合研究中心，西京学院电子信息学院，710123，西安

4 西南技术物理研究所，610041，成都

摘要

量子精密测量利用量子纠缠和量子相干性提高测量精度。在本文中我们简要回顾了在各种复杂系统中的量子精密测量方案，包括非马尔科夫噪声、关联噪声、量子临界系统。另一方面，量子信息的蓬勃发展让我们利用量子模拟实验检验各种理论方案的可行性，并展示复杂系统中丰富的物理现象，例如一维耦合腔阵列中的束缚态、单光子开关和路由器。

关键词　量子精密测量；量子开放系统；非马尔科夫；量子芝诺效应；量子相变；量子模拟

中图分类号　O413　　DOI：10.12202/xxx

## 0 引言

量子精密测量利用量子纠缠和量子相干性提高测量精度[1-2]。以 Ramsey 干涉实验为例，把量子系统初始制备在一个叠加态 $|\psi(0)\rangle = (|a\rangle + |b\rangle)/\sqrt{2}$ 上，然后自由演化到 $|\psi(t)\rangle = (|a\rangle + e^{i\varphi(t)}|b\rangle)/\sqrt{2}$，通过测量相位 $\varphi(t)$ 达到测量频率的目的。为了提高测量精度，一般使用 $n$ 个相同粒子的纠缠态，其测量精度在理论上限正比于 $n^{-1}$，即海森堡极限[2]。一般而言，由于量子系统与环境的相互作用，会导致量子系统发生退相干效应[3]。在纯退相位的马尔科夫噪声作用下，使用最大纠缠态并不会降低测量误差，即标准量子极限 $n^{-1/2}$，即使所需测量时间更短[4]。由于薛定谔方程的幺正性，量子开放系统中初态短时的存活几率总是呈高斯型随时间衰减，此时退相干速率正比于时间，即量子芝诺效应[5-7]。在这样的非马尔科夫噪声中[8]，测量误差正比于 $n^{-3/4}$，即芝诺极限[9]。除了独立热库，在离子阱系统中不同量子比特感受到的噪声之间还存在关联，这启发我们考虑引进辅助量子比特抵消噪声的影响，从而接近无噪声的情况[10]。

众所周知，量子临界系统在相变点附近对参数的变化非常敏感，不仅可以用于解释鸟类迁徙中的导航机制[11]，还可以用于人工导航器件[12]。因此，量子临界现象可用于提高测量精度，其原理在于构造等间距的能谱结构，发生量子相变时能隙会消失，从而导致量子 Fisher 信息发散[13]。但是，双光子耗散由于会保留量子 Rabi 模型的 $Z_2$ 对称性[14-15]，它所具有的非线性会破坏量子临界增强的精密测量的物理基础[16]。

另一方面，最近二十年以来，量子信息与量子计算研究取得了蓬勃发展。量子计算由于量子纠缠与量子相干性可以指数加快量子开放系统严格模拟过程[17-18]。因此，我们提出可以用量子计算机检验各种量子精密测量理论方案的实验可行性[10]，并进行了实验展示[19]。

## 1 量子开放系统与量子精密测量

### 1.1 量子开放系统

由于量子系统与环境的相互作用，系统与环境构成量子开放系统。其量子动力学由哈密顿量来决定，其中环境被描述成无穷多个谐振子模式。以纯退相位噪声为例，其哈密顿量为[3,10]

$$H = H_S + H_B + H_{SB},$$

其中$H_S = \sum_{i=1}^{n} \frac{\omega_0}{2} \sigma_i^z$是系统哈密顿量，$H_B = \sum_k \sum_{i=1}^{n} \omega_k a_{ik}^+ a_{ik}$是环境哈密顿量，$H_{SB} = \sum_k \sum_{i=1}^{n} g_{ik} \sigma_i^z (a_{ik}^+ + a_{ik})$是系统与环境相互哈密顿量。这里，$\omega_0$是量子比特的能级差，$\sigma_i^z$是第$i$个量子比特的 Pauli 矩阵，$a_{ik}^+$（$a_{ik}$）是第$i$个量子比特热库的第$k$个谐振子模式的升（降）算符，$\omega_k$是其本征频率，$g_{ik}$是其与第$i$个量子比特的耦合系数。

量子开放系统的动力学演化一般用 Lindblad 形式量子主方程近似描述[3,10]，即

$$\dot{\rho} = \frac{i}{\hbar}[\rho, H_S] + \sum_{i=1}^{n} \gamma(t) \left( \sigma_i^z \rho \sigma_i^z - \frac{1}{2} \{\sigma_i^z \sigma_i^z, \rho\} \right),$$

其中$\rho$是系统的密度矩阵，$\gamma(t)$是退相位速率，$\{A, \rho\} = A\rho + \rho A$是反对易子。这里，退相位速率是环境关联函数对时间的积分，即

$$\gamma(t) = \int_0^t d\tau g_{ik}^2 a_{ik}^+(\tau) a_{ik}(0).$$

退相位速率对时间的积分即线型函数

$$g(t) = \int_0^t d\tau \gamma(\tau).$$

在推导主方程的过程中[3]，一般需要通过微扰论，将系统与环境相互作用近似到二阶。然后做玻恩近似，假设系统与环境相互作用对环境几乎没有反作用。再做马尔科夫近似，不考虑环境的记忆效应。最后做久期近似，丢掉快变量。而至于量子开放系统的动力学演化的严格计算，采用较多的是级联运动方程[20]，但是其计算复杂度会随着系统的维度和环境关联函数中 e 指数项个数呈指数增加。最近，我们提出了一个量子算法，可以指数加快级联运动方程的严格模拟，且计算的复杂度并不受环境关联函数的复杂度影响[17-18]。其具体形式，我们将在第 4 节进行详述。

1.2 量子精密测量

本文以 Ramsey 干涉实验为例，介绍量子精密测量的基本原理[10]。对于一个 $n$ 量子比特系统，首先将系统制备在最大纠缠态 Greenberger–Horne–Zeilinger（GHZ）态上，即$|\psi(0)\rangle = (|0\rangle^{\otimes n} + |1\rangle^{\otimes n})/\sqrt{2}$。然后让系统在哈密顿量 $H$ 作用下演化一段时间 $t$，最后测量$|0\rangle^{\otimes n}$和$|1\rangle^{\otimes n}$之间的相对相位。

2 海森堡极限与芝诺极限

2.1 无噪声情况

当系统与环境没有相互作用时，即量子闭系统，此时$\gamma(t) = 0$，达到测量精度的海森堡极限，见图 2。

2.2 马尔科夫噪声

当退相位速率不依赖于时间，即马尔科夫噪声，此时$\gamma(t) = C$是一个常数，使用最大纠缠态并不会降低测量误差，但是可以缩短最优测量时间，见图 1。当然，数值计算表明，在马尔科夫环境中使用部分纠缠态，其测量精度相对于无纠缠态可以有适度提高[4]。

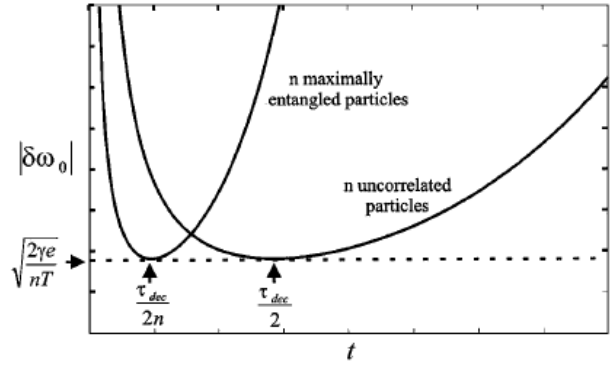

图 1 马尔科夫噪声下，初始态分别为 $n$ 量子比特最大纠缠态与直积态的测量误差动力学[4]

2.3 非马尔科夫噪声

当退相位速率是时间的函数，即非马尔科夫噪声。例如，$\gamma(t) = Ct$，此时测量误差下限即芝诺极限[9]，见图 2。最近，我们采用我们开发的量子算法，通过量子模拟实验在核磁共振量子计算机上，证明量子纠缠态会利用量子芝诺效应提高测量精度 $n^{1/4}$ 倍[19]。由于量子力学是幺正的，量子态短时演化一定是高斯型的，即处于初态的几率会随着时间呈 $t^2$ 减小，此即量子芝诺效应[5,6]。在量子开放系统中，由于频繁测量、以及和环境的相互作用，量子系统处于初态的几率有可能被加速，即量子反芝诺效应，我们证明它是普遍存在的客观现象[5,6]，不会依赖于对量子力学的诠释和理论推导中所做的假设，并通过了超导量子电动力学实验检验[7]。

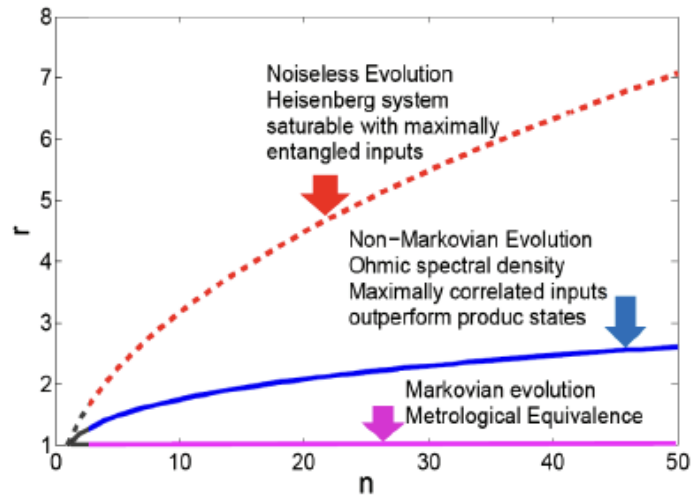

图 2 无噪声、非马尔科夫噪声、马尔科夫噪声下使用最大纠缠态提高的测量精度倍数 $r$ 与量子比特数 $n$ 的关系[9]

3 量子临界增强的精密测量
3.1 动力学框架

2021 年，蔡建明等人提出一种基于量子临界的精密测量的动力学框架[13]。他们发现，在量子 Rabi 模型中，当发生量子相变时，体系的等间距能谱的能隙会消失，此时刻画测量误差的 Fisher 信息会发散，从而导致被测量方差趋近于零。

设系统的哈密顿量为

$$H_\lambda = H_0 + \lambda H_1,$$

其中λ是待测参数。定义$\hat{C} = -i[H_0, H_1]$，$\hat{D} = -i[H_\lambda, \hat{C}]$，$\hat{\Lambda} = i\Delta\hat{C} - \hat{D}$，$|\mu\rangle$是$H_\lambda$的本征态对应本征值$\mu$。如果

$$\Delta\hat{\Lambda} = [H_\lambda, \hat{\Lambda}],$$

可以证明$\hat{\Lambda}|\mu\rangle$ ($\hat{\Lambda}^+|\mu\rangle$) 也是$H_\lambda$的本征态对应本征值$\mu + \Delta$ ($\mu - \Delta$)，即$H_\lambda$的能谱是等间距的。初始时刻，系统被制备在量子态$|\psi\rangle$上，其量子 Fisher 信息随时间的演化为

$$I_\lambda(t) \approx 4\frac{[\sin(\Delta t) - \Delta t]^3}{\Delta^6}[\langle\psi|D^2|\psi\rangle - \langle\psi|D|\psi\rangle^2].$$

当趋近于相变点，且保持$\Delta t$为常数时，$I_\lambda(t)$会发散，此时测量$\lambda$的误差趋近于零。

以量子 Rabi 模型为例，其哈密顿量为

$$H_{\text{Rabi}} = \omega a^+ a + \frac{\Omega}{2}\sigma_z - \lambda(a^+ + a)\sigma_x,$$

其中$a^+$ ($a$) 是频率为$\omega$的单模玻色子的升（降）算符，$\Omega$是二能级原子的能级劈裂，$\lambda$是原子与玻色子的耦合强度，$\sigma_\alpha$ ($\alpha = x, y, z$) 是泡利矩阵。当原子的能级劈裂远大于玻色子频率时，即$\eta = \frac{\Omega}{\omega} \to \infty$，原子与玻色子解耦并为玻色子提供一个有效场，有效哈密顿量表示为

$$H_{\text{np}}^\downarrow = \omega a^+ a + \frac{\omega g^2}{4}(a^+ + a)^2,$$

其中$g = \frac{2\lambda}{\sqrt{\omega\Omega}}$是无量纲的耦合强度。当$g \to 1$时，发生从普通相到超辐射相的量子相变[21]。初始时刻，玻色子被制备在量子态$|\varphi\rangle$上，其量子 Fisher 信息随时间的演化为

$$I_g(t) \approx 16g^2\frac{[\sin(\Delta_g\omega t) - \Delta_g\omega t]^3}{\Delta_g^6}[\langle\varphi|P^4|\varphi\rangle - \langle\varphi|P^2|\varphi\rangle^2],$$

这里$\Delta_g = 2\sqrt{1-g^2}$是无量纲能隙，$P = i(a^+ - a)$是正则动量。当发生量子临界现象时，$I_g(t)$会发散，即测量$g$的误差无穷小。

3.2 噪声敏感性

在本节中，我们考虑各种噪声模型对基于量子临界效应的精密测量方案的影响[He23]。首先，我们考虑在量子开放系统中常见的单光子耗散。在图 3（a）中展示了$F_g(t)$的动力学演化，可见$F_g(t)$做周期性振荡，其振幅先随时间快速增加到极大，然后再缓慢减小。而且当$g$趋近于相变点时，其最大振幅单调增加。尤其值得注意的是，对于不同$g$，$F_g(t)$都是以$\Delta_g\omega$作为频率进行振荡。在图 3（b）中，我们进一步发现，$F_g(t)$的最大值$F_g(t)|_{\max}$与能隙$\Delta_g$之间呈幂律关系$F_g(t)|_{\max} = a\Delta_g^b$，随着$\Delta_g$的减小，即靠近相变点，$F_g(t)|_{\max}$会发散。而且，虽然随着耗散变强，$F_g(t)|_{\max}$会显著减小，对于不同的耗散强度，幂指数都大约是$b = -1.2$。

然后，我们在图 4 中考察温度对$F_g(t)$的效果。在图 4（a）中呈现了不同平均声子数$\bar{n}$时$F_g(t)$的动力学演化，可见$F_g(t)$仍然做周期性振荡，只是随着$\bar{n}$的增加，$F_g(t)$的最大值$F_g(t)|_{\max}$很快减小。在图 4（b）中，我们发现$F_g(t)|_{\max}$与温度之间也满足幂律关系$F_g(t)|_{\max} = aT^b$，其中$b \approx -1$，表明量子临界增强的精密测量在低温时是鲁棒的。

最后，我们通过数值计算考虑双光子耗散下的方案鲁棒性，在图 5（b）中，我们发现，$F_g(t)$的最大值$F_g(t)|_{\max}$与能隙$\Delta_g$之间虽然仍然满足幂律关系$F_g(t)|_{\max} = a\Delta_g^b$，但是随着$\Delta_g$的减小，$F_g(t)|_{\max}$会减小，即越靠近相变点测量的精度越低。这点也可以从在图 5（a）中找到原因，对于不同$g$，$F_g(t)$不再以$\Delta_g\omega$作为频率进行振荡，即双光子耗散破坏了体系能谱的等间距结构。

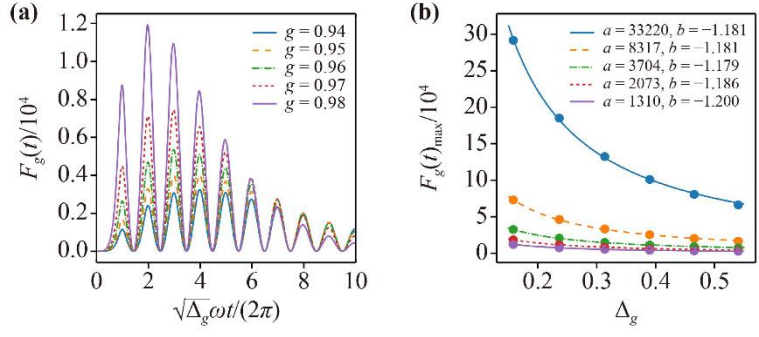

图 3 发生单光子耗散时，（a）不同耦合强度下，$g$方差的倒数$F_g(t)$随时间的演化；（b）$F_g(t)$最大值与能隙关系[16]。

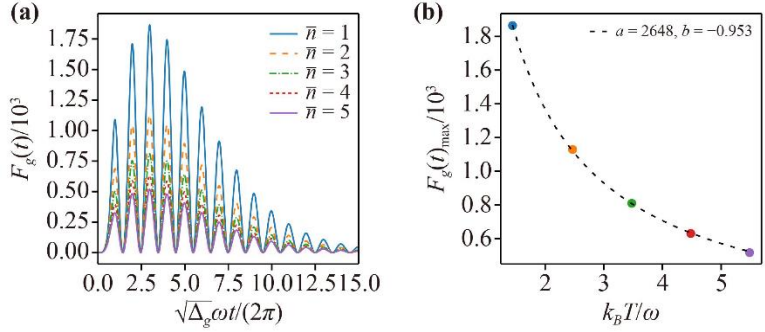

图 4 发生单光子耗散时，（a）不同温度下，$g$方差的倒数$F_g(t)$随时间的演化；（b）$F_g(t)$最大值与温度关系[16]。

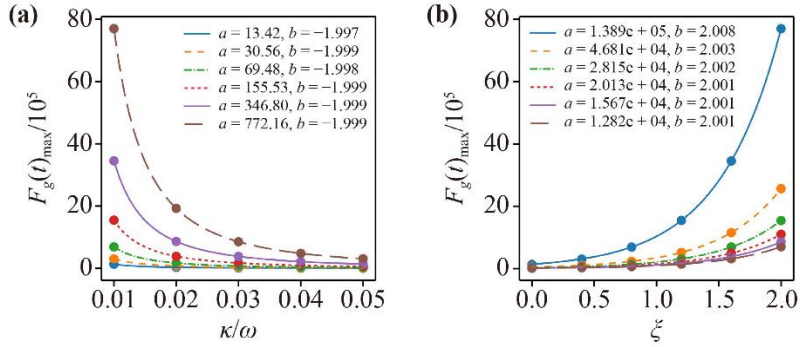

图 5 发生双光子耗散时，（a）不同耦合强度下，$g$方差的倒数$F_g(t)$随时间的演化；（b）$F_g(t)$最大值与能隙关系[16]。

4 量子模拟

4.1 量子模拟方法

量子模拟一般需要生成大量具有相同初态的样本，然后在不同哈密顿量$H_Q = H_{QS} + H_{QN}$作用下进行时间演化，然后对系综取平均得到的密度矩阵符合级联运动方程。此处，$H_{QS}$模拟量子开放系统中$H_S$，$H_{QN}$模拟环境的作用即$H_B + H_{SB}$，其具体形式如下

$$H_{QN} = \sum_m \beta_m(t)|m\rangle\langle m|,$$

其中

$$\beta_m(t) = \sum_{j=1}^{N_c} \alpha_m F(\omega_j)\omega_j \cos(\omega_j t + \phi_j^{(m)}).$$

这里，$\alpha_m$是噪声的幅度，$F(\omega_j)$刻画噪声的关联函数形式，$\omega_j = j\omega_0$，$\omega_0$是基频，$N_c\omega_0$是截止频率，$\phi_j^{(m)}$是一组均匀分布在$[0,2\pi)$的随机数。噪声的关联函数是$S_m(t) = \langle \beta_m(t+\tau)\beta_m(t)\rangle$，即

$$S_m(t) = \lim_{T\to\infty}\int_{-T}^{T} dt\, \beta_m(t+\tau)\beta_m(t) = \left(\frac{\alpha_m}{2}\right)^2 \sum_j [F(\omega_j)\omega_j]^2 (e^{i\omega_j\tau} + e^{-i\omega_j\tau}),$$

它与$t$无关，而只依赖于$\tau$。对它做傅里叶变换，得到噪声的功率谱密度

$$\tilde{S}_m(\omega) = \int_{-\infty}^{\infty} d\tau\, e^{-i\omega\tau}\langle\beta_m(t+\tau)\beta_m(t)\rangle = \pi\frac{\alpha_m^2}{2}\sum_j [F(\omega_j)\omega_j]^2 (\delta(\omega-\omega_j) + \delta(\omega+\omega_j)).$$

这里模拟噪声效果采用的是热库调控（bath engineering）技术，最早在离子阱系统中提出、并模拟了各种谱密度形式的退相位和弛豫噪声[22]。要让模拟的动力学跟目标量子开放系统动力学完全一致，需要让两者的哈密顿量$H_S$和$H_{QS}$、以及两者的关联函数$C_m(t)$和$S_m(t)$分别相同。当然，在实际模拟中，由于被模拟的量子开放系统和系综的能量尺度不一致，两者之间一般差一个比例系数，即

$$\frac{H_S}{H_{QS}} = \frac{C_m(t)}{S_m(t)}.$$

在量子模拟实验中，实际操作一般分为以下步骤：首先，通过以上关系找到符合条件的$H_{QS}$和$S_m(t)$。因为$S_m(t)$确定，噪声哈密顿量$H_{QN}$也就确定了。然后，再通过量子—经典杂化版 gradient ascent pulse engineering (GRAPE)算法，找到实验中可行的脉冲序列[23-24]，最后在实验中实现。我们已经用该算法在核磁共振量子计算机中模拟了光合作用中能量传输和电荷分离[17,25]，验证了我们提出的应该用全局和局域变量度量量子开放系统动力学的非马尔科夫性[8]，并检验了量子精密测量方案的实验可行性，见图 7[10,19]。此外，通过理论分析，证明在超导量子计算机中也是切实可行的[26]。

4.2 检验量子精密测量方案

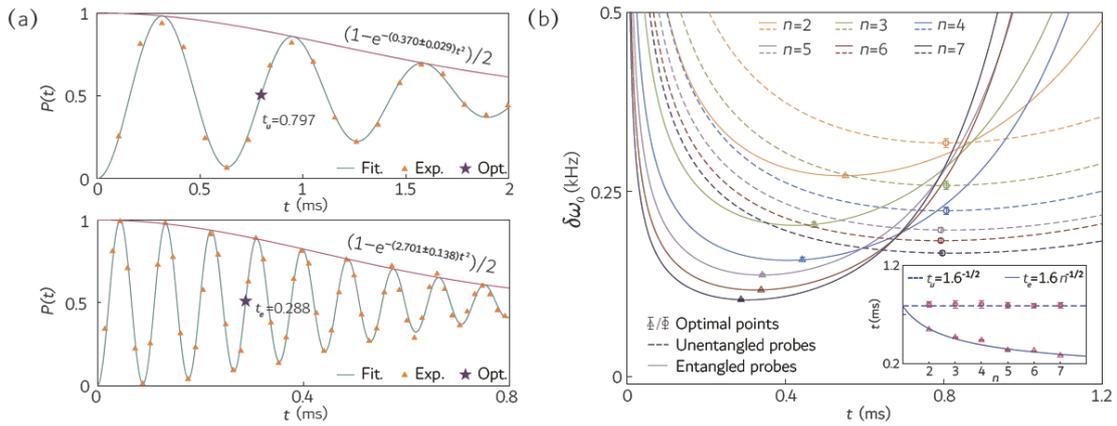

图 6 量子模拟（a）马尔科夫噪声，（b）非马尔科夫噪声中的精密测量[10].

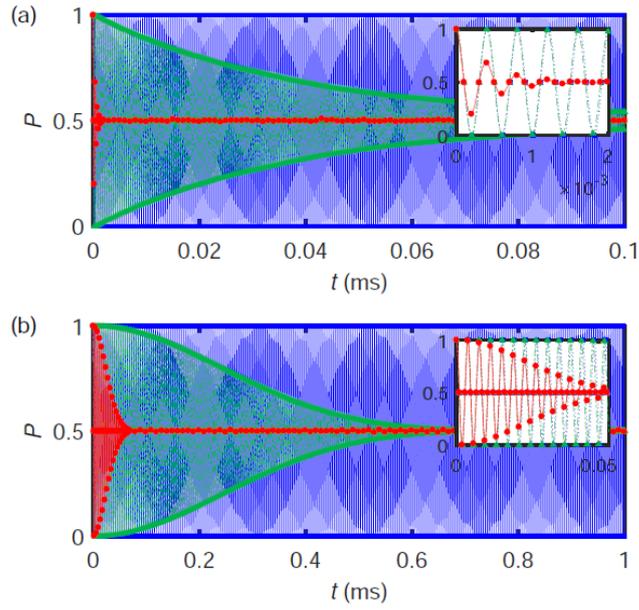

图 7 （a）上：直积态量子动力学；下：GHZ 态量子动力学；（b）测量误差动力学[19].

我们的方法对于检验一类量子精密测量方案都有效。例如，2019 年兰州大学安俊鸿等人提出，当系统与环境相互作用满足特定条件，存在束缚态时，可以达到无噪声的海森堡极限[27]。2010 年，我们在研究一维耦合腔阵列与一个二能级原子相互作用时，通过严格求解量子开放系统动力学发现存在束缚态[28]。这个模型在数学上等价于一个二能级系统与一个能带耦合。因此，如果我们在量子计算机上构造一个具有相同能谱结构的量子系统，就可以演示通过束缚态达到量子精密测量的海森堡极限。当然，实现该方案，还有其它办法。如果在量子计算机上对一个量子比特施加纵向弛豫噪声，其能谱具有一维耦合腔阵列相同的能带结构，与前一种方法相比较，后者更节省资源，但是前者可以模拟耦合腔阵列中丰富的物理现象，例如单光子开关、单光子量子路由器、单原子超腔、双原子超关联辐射特性和巨原子[29]等等。

除了以上方案，我们的量子模拟方法还可以用于检验量子临界增强的精密测量方案。2021 年，蔡建明等人提出一种基于量子临界的精密测量的动力学框架[13]。他们发现，在量子 Rabi 模型中，当发生量子相变时，体系的等间距能谱的能隙会消失，此时刻画测量误差的 Fisher 信息会发散，从而导致被测量方差趋近于零。为了模拟该方案，我们可以在 $N$ 量子比

特系统中构造等间距能谱，当能级间距消失时，即等效于发生量子临界现象。为了有效模拟热力学极限，我们可以逐渐增大 $N$，如果此时测量误差趋近于 0，即方案得证。

5 结论

在本文中，我们简要回顾了各种复杂系统中的精密测量方案。对于一般情况，即马尔科夫噪声，使用最大纠缠态并不会提高测量精度[4]。当环境是非马尔科夫噪声时，使用最大纠缠态可以达到芝诺极限[9,30]。在离子阱系统中，由于不同量子比特感受到的噪声之间存在关联，可以使用辅助量子比特，并将系统制备在合适的初态上，从而达到没有噪声的海森堡极限[10]。当量子开放系统存在束缚态时，令人吃惊的是可以达到没有噪声的海森堡极限[27]。对于 Rabi 模型和 LMG 模型一类具有等间距能谱的量子临界系统，当发生单光子耗散时，趋近于量子相变点，量子 Fisher 信息和测量精度会发散[13]；当发生双光子耗散时，量子临界现象反而会增加测量误差[10]。

我们首次提出可以应用量子模拟实验，检验精密测量方案的可行性[10]。并证明最大纠缠态确实可以用量子芝诺效应提高测量精度 $n^{1/4}$ 倍[19]。我们可以通过构造有结构的热库，来验证束缚态是否可以达到海森堡极限[27]；或者构造具有等间距能谱的量子临界系统，来展示量子临界增强的精密测量方案。除了验证精密测量方案，我们的量子模拟算法，还可以用来演示一维耦合腔阵列中丰富的物理现象。

Quantum metrology in complex systems and experimental verification by quantum simulation
Ai Qing[1]    Wang YangYang[2,3]    Qiu Jing[4]
(1 The Department of Physics, Beijing Normal University, 100875, Beijing, China)
(2 Shaanxi Engineering Research Center of Controllable Neutron Source, School of Electronic Information, Xijing University, 710123, Xi'an, China)
(3 Shaanxi International Joint Research Center for Applied Technology of Controllable Neutron Source, School of Electronic Information, Xijing University, 710123,
Xi'an, China)
(4 Southwest Institute of Technical Physics, 610041, Chengdu, China,)

**Abstract**    Quantum metrology based on quantum entanglement and quantum coherence improves the accuracy of measurement. In this paper, we briefly review the schemes of quantum metrology in various complex systems, including non-Markovian noise, correlated noise, quantum critical system. On the other hand, the booming development of quantum information allows us to utilize quantum simulation experiments to test the feasibility of various theoretical schemes and demonstrate the rich physical phenomena in complex systems, such as bound states in one-dimensional coupled cavity arrays, single-photon switches and routers.
**Keywords**    Quantum metrology; Open quantum system; Non-Markovian; Quantum Zeno effect; Quantum phase transition; Quantum simulation